\documentclass[twocolumn,epjc3]{svjour3}
\smartqed 
\RequirePackage{graphicx} 
\RequirePackage[numbers,sort&compress]{natbib}
\journalname{Eur. Phys. J. C}
\usepackage{amsmath,latexsym,epsfig,rotate,amssymb}

\newcommand{\vslash}[1]{#1 \hspace{-0.5 em} /}

\begin{document}
\title{Angular distribution and forward--backward asymmetry of the Higgs-boson decay to photon and lepton pair }
\author{Alexander Yu. Korchin\thanksref{addr1,addr2,e1}        \and
        Vladimir A. Kovalchuk\thanksref{addr1,e2}}
\thankstext{e1}{e-mail: korchin@kipt.kharkov.ua}
\thankstext{e2}{e-mail: koval@kipt.kharkov.ua}
\institute{NSC `Kharkov Institute of Physics and Technology',
61108 Kharkiv, Ukraine\label{addr1}          \and
          V.N.~Karazin Kharkiv National University, 61022 Kharkiv, Ukraine\label{addr2}
       }
       \date{Received: date / Accepted: date}
\maketitle

\begin{abstract}

The Higgs-boson decay $h \to \gamma \ell^+ \ell^-$ for various lepton states $\ell = (e, \, \mu, \, \tau)$ is analyzed. The differential decay width and forward--backward asymmetry are calculated as functions of the dilepton invariant mass in a model where the Higgs boson interacts with leptons and quarks via a mixture of scalar and pseudoscalar couplings. These couplings are partly constrained from data on the  decays to leptons, $h \to \ell^+ \ell^-$, and quarks $h \to q \bar{q} $ (where $q = (c, \, b)$), while the Higgs couplings to the top quark are chosen from the two-photon and two-gluon decay rates. Nonzero values of the forward--backward asymmetry will manifest effects of new physics in the Higgs sector. The decay width and asymmetry integrated over the dilepton invariant mass are also presented.


\end{abstract}

\maketitle

\section{\label{sec:Introduction}Introduction}

Since the discovery of the Higgs boson~\cite{ATLAS:2012,CMS:2012}
its decay channels have been extensively studied. In general, the
decay pattern and properties of the $h$ boson are
consistent~\cite{CMS:2012sp} with the quantum numbers $J^{PC} =
0^{++}$ of the boson in the standard model (SM). Yet the nature of
$h$ needs to be clarified and will be investigated in detail in
the next run of the LHC after its upgrade.

In many extensions of the SM a more complicated Higgs sector can
exist, and some of the Higgs bosons may not have definite $CP$
parity~\cite{Pilaftsis:1999np,Barger:2009pr,Branco:2012pre}. This
aspect of the Higgs-boson physics is important for clarification
of the origin of the $CP$ violation, and possible additional
mechanisms beyond the $CP$ violation via the CKM matrix which can
contribute to the observed matter--antimatter asymmetry in the
Universe~\cite{Bailin}.

The $CP$ properties of the Higgs boson were addressed
for the two-photon decay $h \to \gamma \, \gamma$ in a model with vectorlike fermions~\cite{Voloshin:2012}. It was shown that the mutual orientation of linear polarizations of the photons carries information on the $CP$ violation.
This idea was elaborated in Ref.~\cite{Bishara:2013vya}, where the Bethe--Heitler conversion on nuclei of the two photons to electron--positron pairs was suggested as a means to probe the $CP$ violation in the Higgs coupling to photons.
In Ref.~\cite{Kovalchuk:2008} the author analyzed possibilities of
observation of the $CP$ violation effects in the Higgs decays $ h
\to V_1 \, V_2 \, \to (f_1 \, \bar{f}_2) \, (f_3 \, \bar{f}_4)$ to
various final lepton and quark pairs, where $V = W^\pm, \, Z$.

In Refs.~\cite{Korchin:2013,Korchin:2013jja} the authors
suggested to study the $CP$ violation effects in the Higgs sector
in the decay $h \to \gamma Z$ via the polarization parameters of
the photon, or $Z$ boson. A direct way for this
is the forward--backward (FB) asymmetry in the decay $h \to \gamma
Z \to \gamma f \bar{f}$, where the $Z$ boson on
the mass shell decays to fermions. This observable
vanishes in the SM and therefore it carries information on physics
beyond the SM. Estimates of $CP$ violation effects in some models
of new physics were made in \cite{Korchin:2013,Korchin:2013jja}.

The invariant-mass distributions in the Higgs decay to the $\gamma \ell^+ \ell^- $ and $\gamma q \bar{q}$ \ final states were intensively explored in Refs.~\cite{Abbasabadi:1995,Abbasabadi:1997,Abbasabadi:2000,LBChen:2012,Dicus:2013,Passarino:2013,Dicus:2013a,Dicus:2014}.
The first experimental study of the process $h \to \gamma \mu^+ \mu^- $ by the CMS collaboration was recently reported in~\cite{CMS:2014eeg}. The analysis in~\cite{CMS:2014eeg} was performed for dimuon invariant mass less than 20 GeV.

In Refs.~\cite{Sun:2013,Akbar:2014} the authors discussed the angular distribution of the leptons $\ell = (e, \, \mu, \, \tau)$ in the decay $h \to \gamma \ell^+ \ell^-$ in the framework of the SM. The importance of the FB asymmetry was emphasized, and its nonzero values were found. At the same time, beyond the SM, in Ref.~\cite{Chen:2014} the FB asymmetry was proposed as a probe for $CP$-violating Higgs coupling to $Z \gamma$ and $\gamma \gamma$ states.

The FB asymmetry sure enough is an informative observable
which can be of interest for future experiments at the LHC.
In the present paper we address the decay $h \to \gamma \ell^+ \ell^-$ in
some detail. In addition to the loop mechanism $h \to \gamma Z^*
\to \gamma \ell^+ \ell^-$ considered
in~\cite{Korchin:2013,Korchin:2013jja}, we include here the photon
bremsstrahlung off leptons, i.e. tree-level amplitudes for $h \to \gamma \ell^+ \ell^- $, and the loop amplitude $h \to \gamma \gamma^* \to
\gamma \ell^+ \ell^-$. We remark that in the framework of the
SM the FB asymmetry is equal to zero as a consequence of the
scalar nature of the Higgs boson. This asymmetry can take
nonzero values only in models beyond the SM and therefore this observable is sensitive to possible $CP$ violation in the Higgs sector.

To estimate values of this asymmetry we apply a model in which the Higgs boson couples to fermions with a mixture
of the scalar (S) and pseudoscalar (PS) interactions. The strength
of the S and PS couplings, $1 + s_f$ and $p_f$, respectively, are partly constrained from the LHC measurements of the decay rates $h \to
\ell^+ \ell^- $ and $h \to q \bar{q}$ \ (where $q=(c, \, b)$)~\cite{CMS:2014,ATLAS:2014}. As for the Higgs interaction with the top quark, the corresponding couplings are chosen from experimental information on the two-photon, $h \to \gamma \gamma$, and two-gluon, $h \to g g$, decay widths.

In this model, for the decays $h \to \gamma
\ell^+ \ell^-$ we derive the distribution over the angle $\theta$ between the momentum of the lepton (in the rest frame of the pair $\ell^+ \ell^-$) and momentum of the photon (in the rest frame of $h$). The presence of the PS $h f \bar{f}$ coupling gives rise to the linear in $\cos \theta$ terms in this distribution, and thereby to a FB asymmetry. We calculate the differential decay width and FB asymmetry as
functions of the dilepton invariant mass squared $q^2 = (q_+ +
q_-)^2$ ($q_+$ and $q_-$ are the four-momenta of leptons). The widths and FB asymmetries integrated over the invariant mass are also discussed.

The paper is organized as follows. In Sect.~\ref{sec:formalism}
amplitudes and angular distribution in $h \to \gamma \ell^+
\ell^-$ are presented. The loop contributions are defined for the
S and PS Higgs couplings to the fermions. The FB asymmetry is
discussed. In Sect.~\ref{sec:results} the differential decay width and FB
asymmetry for various leptons are calculated. The results of
the calculation are discussed. Section~\ref{sec:conclusions} contains the
conclusions. In~\ref{app:A} the loop integrals are
defined, and in~\ref{app:B} vanishing of
the contribution from axial-vector $Z f \bar{f}$ coupling to the fermion-loop
diagrams is shown.


\section{\label{sec:formalism} Formalism  }

\subsection{Amplitudes and angular distribution}

There are models with more than one Higgs doublet which induce
$CP$ violation due to the specific coupling of neutral Higgs
bosons to fermions. We assume that the couplings of $h$ boson to the fermion
fields, $\psi_f$, are given by the Lagrangian including both
scalar and pseudoscalar parts,
\begin{equation}\label{eq:001}
{\cal L}_{hff}=-\sum_{f = \ell, \, q} \frac{m_f}{v}\,h\,{\bar
\psi_f}\left(1+s_f+i\,p_f \gamma_5\right)\psi_f \,,
\end{equation}
where $v=\left(\sqrt{2}G_{\rm F}\right)^{-1/2}\approx 246$ GeV is
the vacuum expectation value of the Higgs field,  $G_F = 1.166378
\times 10^{-5}$  GeV$^{-2}$ is the Fermi constant~\cite{PDG:2012},
$m_f$ is the fermion mass and $s_f$, $p_f$ are real parameters
($s_f=p_f=0$ corresponds to the SM). Equation~(\ref{eq:001}) can be
considered as a phenomenological parametrization of effects of new
physics. As for the Higgs interaction with the $W^\pm$ and $Z$ bosons, it is assumed to be the same as in the SM.

We consider the decay of the zero-spin Higgs $h$ boson
\begin{equation}\label{eq:002}
h(p) \, \to \, \gamma (k, \, \epsilon(k) ) +\ell^+ (q_+) + \ell^- (q_-) \,,
\end{equation}
where the four-momenta of the $h$ boson, photon, and leptons are
$p$, \ $k$, \  $q_+$, \ $q_-$ respectively, and $\epsilon (k)$ is
the polarization four-vector of the photon.

The differential decay width can be written as
\begin{equation}
\frac{ {\rm d} \Gamma }{{\rm d}  q^2 \, {\rm d} \cos\theta }=
\frac{\beta_{\ell}  (m_h^2 -q^2)}{(8 \pi)^3 \, m_h^3} \, |{\cal M}|^2 \,,
\label{eq:003}
\end{equation}
where $m_h$ is the mass of the $h$ boson, $q \equiv q_+ + q_- $, \
$q^2$ is the invariant mass squared of the lepton pair,
$\beta_{\ell} = \sqrt{1-4 m_{\ell}^2/q^2}$ is the lepton velocity
in the rest frame of the lepton pair.  The polar angle $\theta$ is
defined in this frame and it is the angle between the momentum of lepton $l^+$
and the axis opposite to the direction of the Higgs-boson momentum.

\begin{figure}[tbh]
\begin{center}
\includegraphics[width=0.40\textwidth]{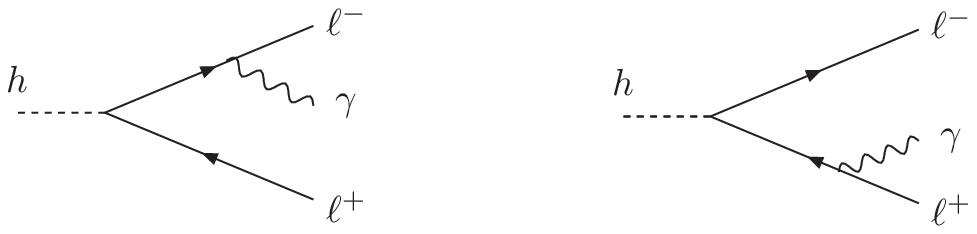}

\includegraphics[width=0.47\textwidth]{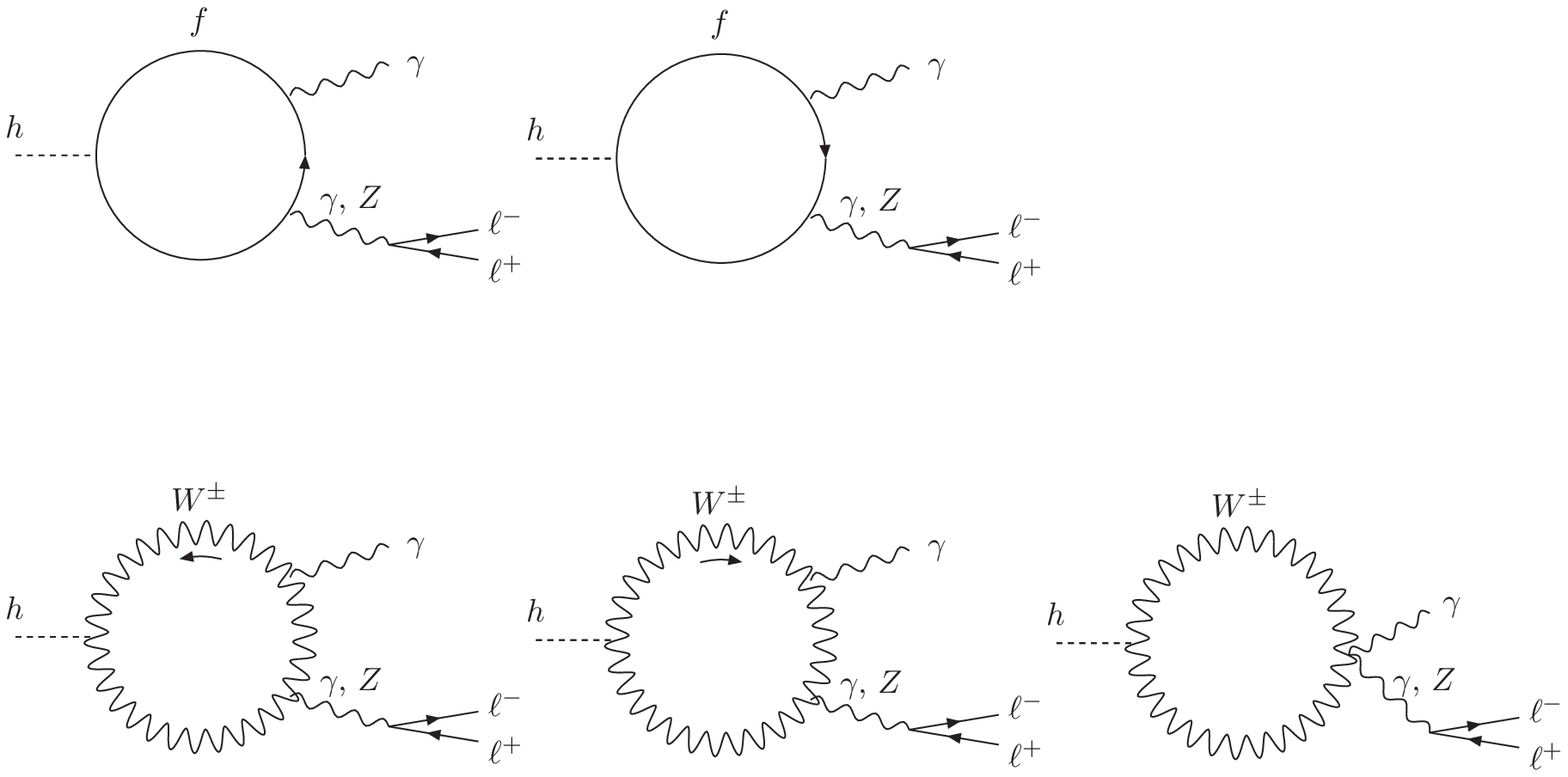}
\end{center}
\caption{Diagrams for the process $h \to \gamma \ell^+ \ell^-$.
The {\it upper row} shows the tree-level (bremsstrahlung) amplitudes,
and the loop diagrams are drawn below. Fermions $f$
are indicated by the {\it solid lines}, gauge bosons $W^\pm, \, Z, \,
\gamma$ by the {\it wavy lines}, and $h$ boson by the {\it dashed
lines} }
\label{fig:diagrams}
\end{figure}

The amplitude of the decay is
\begin{equation}\label{eq:004}
{\cal M} \, = \, {\cal M}_{tree} \, + \, {\cal M}_{loop}\, ,
\end{equation}
where the tree-level amplitude (Fig.~\ref{fig:diagrams}) is
\begin{eqnarray}
{\cal M}_{tree} &=& c_0 \, \epsilon_\mu^* (k) \, \bar{u}(q_-) (1+s_{\ell}+i \,p_{\ell} \gamma_5 ) \nonumber  \\
& \times &  \Bigl( \frac{2q_+^\mu + \vslash{k} \gamma^\mu}{2 k \cdot q_+ } - \frac{2q_-^\mu + \gamma^\mu \vslash{k}}{2 k \cdot q_- }
\Bigr) \, v(q_+)  \, ,
\label{eq:005}
\end{eqnarray}
where
\begin{equation}
c_0 = e m_{\ell} Q_{\ell} \, ({\sqrt{2} G_F})^{1/2}\, ,
\label{eq:c_0}
\end{equation}
$e = \sqrt{4 \pi \alpha_{G_F}}$ \ is the positron charge,
$Q_{\ell} = -1$ (lepton charge in units of $e$) and  $m_{\ell}$ is
the lepton mass. The electromagnetic coupling in the
$G_F$-scheme~\cite{Heinemeyer:2013} is \ $\alpha_{G_F} =
\sqrt{2}G_F m_W^2(1-m_W^2 /m_Z^2)/ \pi$, where $m_W \, (m_Z)$ is
the mass of the $W$ ($Z$) boson.
For the rest we follow the standard definition of the $\gamma$ matrices and
lepton spinors (see, e.g. \cite{Pes95}).

The loop contributions $h \to \gamma \, \gamma^* / Z^* \to \gamma \ell^+ \ell^-$
(see Fig.~\ref{fig:diagrams}) can be written in the form
\begin{eqnarray}
 {\cal M}_{loop} &=& \epsilon_\mu^* (k) \, [ \, (q^\mu k^\nu -  g^{\mu \nu} k\cdot q)   \nonumber \\
& \times &  \bar{u}(q_-) ( c_1 \gamma_\nu + c_2 \gamma_\nu \gamma_5) v(q_+)  \nonumber
\\
&-& \epsilon^{\mu \nu \alpha \beta} k_\alpha q_\beta \, \bar{u}(q_-) ( c_3 \gamma_\nu + c_4 \gamma_\nu \gamma_5 ) v(q_+) \,  ] \, ,
\label{eq:006}
\end{eqnarray}
with coefficients $c_1, \ldots , c_4 $ which are
specified below in terms of the loop functions,  and
$\epsilon_{0123} = +1$.  Here we follow notation in
Refs.~\cite{Sun:2013,Akbar:2014}.

Note that we do not take into account the loop contributions of the type
$h \to \gamma \ell^+ \ell^- $ (the so-called box diagrams). The contribution of these diagrams to the considered decay is negligibly small in the
SM~\cite{Abbasabadi:1995,Abbasabadi:1997}.
Also the processes $h \to \gamma V \to \gamma \ell^+ \ell^-$, where $V$ is intermediate vector resonance decaying into the $\ell^+ \ell^-$ pair, can contribute to the decay $h \to \gamma \ell^+ \ell^-$.
In particular, resonant production of the quarkonium states $J/ \psi \, (c \bar{c})$ and $\Upsilon (1S) \, (b \bar{b})$ is of interest for studying the $h q \bar{q}$ coupling (see, for example \cite{Bodwin:2013,Kagan:2014,Gao:2014,London:2014}).
The account of such mechanisms lies beyond the scope of the present work.

We evaluate the amplitude (\ref{eq:004}) squared, sum over the lepton and
photon polarizations, and obtain the following result in the model (\ref{eq:001})
\begin{eqnarray}\label{eq:007}
|{\cal M}|^2  &=&  c_0^2 \, \bigl[  (1+s_{\ell})^2 \, A + p_{\ell}^2 \, \widetilde{A}  \bigr] \nonumber \\
&& + 2 \, c_0 \, \bigl[  (1+s_{\ell}) \, {\rm Re} (c_1) \, B + p_{\ell} \, {\rm Im} (c_2) \, \widetilde{B} \, \nonumber \\
&& + (1+s_{\ell}) \, {\rm Im} (c_4) \, C + p_{\ell} \,  {\rm Re}(c_3) \, \widetilde{C} \bigr] \nonumber \\
&& + \bigl( |c_1|^2 + |c_3|^2  \bigr) \, D  + \bigl( |c_2|^2 + |c_4|^2  \bigr) \, E  \nonumber \\
&& + 2 \, {\rm Im} \bigl(c_1 c_4^* + c_2 c_3^*\bigr) \, F \, .
\end{eqnarray}
The fact that $c_0$ is real while $c_1, \ldots, c_4$
are generally complex-valued is used in derivation of (\ref{eq:007}).

The coefficients in Eq.~(\ref{eq:007}) are defined as follows
(we use below the notation $z \equiv \cos \theta$)
\begin{eqnarray}
A&=& \frac{16}{(1- \beta_{\ell}^2 z^2 )^2 (m_h^2-q^2)^2}
[\, (m_h^4 + q^4  \nonumber \\
&& - 8 m_{\ell}^2 q^2)(1-\beta_{\ell}^2 z^2) +32 m_{\ell}^4 - 8 m_h^2 m_{\ell}^2 \, ]\, , \label{eq:008} \\
\widetilde{A} &=& \frac{16}{(1- \beta_{\ell}^2 z^2 )^2 (m_h^2-q^2)^2}
[\, (m_h^4+q^4) \nonumber   \\
&& \times \, (1-\beta_{\ell}^2 z^2)
 -  8 m_h^2 m_{\ell}^2\, ]\, , \label{eq:009}
\end{eqnarray}
\begin{eqnarray}
B &=& - \frac{8 m_{\ell} }{(1- \beta_{\ell}^2 z^2 )}  [\, m_h^2 -q^2 +q^2 \beta_{\ell}^2 (1-z^2 )\, ] \, , \label{eq:010} \\
\widetilde{B} &=&  - \frac{8 m_{\ell}  }{(1- \beta_{\ell}^2 z^2 )} \,  (m_h^2-q^2) \, \beta_{\ell} \, z \, ,  \label{eq:011}\\
C & =& - \frac{8 m_{\ell} }{(1- \beta_{\ell}^2 z^2 )} \, (m_h^2-q^2) \, \beta_{\ell} \, z \, , \label{eq:012} \\
\widetilde{C} &=& \frac{8 m_{\ell} }{(1- \beta_{\ell}^2 z^2 )} (m_h^2-q^2)\, , \label{eq:013} \\
D &=& \frac{1}{2} (m_h^2-q^2)^2 \, [\, q^2 (1 + \beta_{\ell}^2 z^2  ) +4 m_{\ell}^2 \,]\, , \label{eq:014} \\
E &=&  \frac{1}{2} (m_h^2-q^2)^2 \, q^2 \, \beta_{\ell}^2 \, (1 + z^2 )\, ,  \label{eq:015} \\
F &=& - (m_h^2-q^2)^2 \, q^2 \, \beta_{\ell} \,  z \,. \label{eq:016}
\end{eqnarray}

The FB asymmetry is defined as (see, e.g. \cite{Korchin:2013, Korchin:2013jja} and ~\cite{Sun:2013,Akbar:2014})
\begin{equation}\label{eq:017}
A_{\rm FB} (q^2) =\Big( {\frac{{\rm d} \Gamma_F }{{\rm d} q^2} \, - \, \frac{{\rm d} \Gamma_B }{{\rm d} q^2}} \Big) {\Big( \frac{{\rm d} \Gamma_F }{{\rm d} q^2} \,+ \, \frac{{\rm d} \Gamma_B }{{\rm d} q^2} \Big)^{-1} }\,,
\end{equation}
where
\begin{eqnarray}
\frac{{\rm d} \Gamma_F }{{\rm d} q^2} & \equiv &
\int_0^1 \frac{{\rm d} \Gamma}{{\rm d} q^2 \, {\rm d} \cos\theta}\, {\rm d}  \cos\theta \,, \nonumber  \\
\frac{{\rm d} \Gamma_B }{{\rm d} q^2} &\equiv &
\int_{-1}^0 \frac{ {\rm d} \Gamma}{ {\rm d} q^2 \, {\rm d} \cos\theta}\, {\rm d}  \cos\theta \, .
\, \label{eq:018}
\end{eqnarray}

As only the coefficients $\widetilde{B}, \, C$ and $F$ are linear
in $\cos \theta$, then it is seen from
Eqs.~(\ref{eq:008})-(\ref{eq:016}) that the numerator of the
asymmetry  (\ref{eq:017}) is determined by the imaginary part of
the terms $c_2$, $c_4$ and the combination $c_1 c_4^* + c_2
c_3^*$:
\begin{eqnarray}
&& \frac{{\rm d} \Gamma_F }{{\rm d} q^2} - \frac{{\rm d} \Gamma_B }{{\rm d} q^2} =
-\frac{2  (m_h^2-q^2)^2 }{(8 \pi)^3 \, m_h^3} \nonumber \\
&& \times \Big[ c_0  \, \Big( p_\ell \, {\rm Im}(c_2) + (1+s_\ell) \, {\rm Im} (c_4) \Big) \, 8 m_\ell \ln \Big( \frac{q^2}{4m_\ell^2} \Big)  \nonumber \\
&& + \, {\rm Im} \big(c_1 c_4^* + c_2 c_3^*\big) \, (q^2  - 4 m_\ell^2) \, (m_h^2-q^2)  \Big]\, .
\label{eq:delta_Gamma}
\end{eqnarray}

It may be instructive to analyze the asymmetry (\ref{eq:017}) in
the limit of zero lepton masses. Putting $m_{\ell} =0$ in
(\ref{eq:007}),  (\ref{eq:010})--(\ref{eq:016}) and
(\ref{eq:delta_Gamma}) one obtains the distribution over the
$\ell^+ \ell^-$ invariant mass
\begin{equation}
 \frac{{\rm d} \Gamma }{{\rm d} q^2} \equiv   \frac{{\rm d} \Gamma_F }{{\rm d} q^2} +  \frac{{\rm d}
\Gamma_B }{{\rm d} q^2}
 = \frac{ (m_h^2 -q^2)^3 \, q^2 }{6 \, (4 \pi)^3 \, m_h^3 } \,
 \sum_{j=1}^4 \, |c_j|^2 \,
\label{eq:Gamma_q^2}
\end{equation}
and the FB asymmetry
\begin{equation}
A_{\rm FB}(q^2) = -\frac{3}{2} \frac{{\rm Im}\big(c_1 c_4^* +c_2 c_3^*\big)}{|c_1|^2 + |c_2|^2 + |c_3|^2 +  |c_4|^2} \,.
\label{eq:FB-asymmetry_m=0}
\end{equation}

For further reference we also introduce the integrated over $q^2$ asymmetry~\cite{Sun:2013}:
\begin{equation}
\langle A_{\rm FB} \rangle = \int \Big( {\frac{{\rm d} \Gamma_F }{{\rm d} q^2} \, - \, \frac{{\rm d} \Gamma_B }{{\rm d} q^2}} \Big) \, {\rm d} q^2  \,
\Big (\int
\frac{{\rm d} \Gamma }{{\rm d} q^2} {\rm d} q^2  \Big)^{-1}   \,,
\label{eq:AFB_int}
\end{equation}
for appropriate integration limits $q^2_{min} \geq 4 m_\ell^2 $ and $q^2_{max} \le m_h^2$.


\subsection{Loop contributions}

Let us specify the loop contributions in Fig.~{\ref{fig:diagrams}}
to the coefficients $c_1, \ldots , c_4$. We introduce below the
Weinberg angle $\theta_W$ and the notation $s_W \equiv \sin \theta_W$
and $c_W \equiv \cos \theta_W$.

We evaluate the loop diagrams using the Lagrangian (\ref{eq:001})
for the $h f \bar{f}$ vertex. The scalar coupling of the Higgs to
fermions contributes to the coefficients $c_1, \, c_2$ which read
\begin{eqnarray}
c_1 &=&  \frac{1}{2} \,  \frac{g_{V,  \ell}}{q^2 - m_Z^2 + i m_Z \Gamma_Z} \,
\Pi_{ Z} + \frac{Q_{\ell}}{q^2} \,  \Pi_{\gamma } \, , \label{eq:c_1} \\
c_2 &=&  - \frac{1}{2} \, \frac{g_{A,  \ell}}{q^2 - m_Z^2 + i m_Z \Gamma_Z} \, \Pi_{ Z} \, , \label{eq:c_2}
\end{eqnarray}
where $m_Z$  ($\Gamma_Z$) is the mass (total decay width) of the $Z$ boson,  and
\begin{eqnarray}
\Pi_{ Z} &=&   \frac{e g^3}{16 \pi^2 m_W } \,
\bigl[ (1+s_f) \, \frac{2 g_{V,  f}}{c_W^2} \, N_f Q_f A_f (\lambda_f^\prime, \lambda_f)   \nonumber \\
&&  +  A_W (\lambda_W^\prime, \lambda_W) \bigr] \, , \label{eq:Pi_Z} \\
\Pi_{\gamma} &=&   \frac{e^3 g}{16 \pi^2 m_W} \,
\bigl[ (1+s_f) \, 4 Q_f^2 N_f  \, A_f (\lambda_f^\prime, \lambda_f)  \nonumber \\
&& + A_W (\lambda_W^\prime, \lambda_W) \bigr] \, .   \label{eq:Pi_gamma}
\end{eqnarray}
Here $g = 2 m_W (\sqrt{2}G_F)^{1/2}$ is the $SU(2)_L$ coupling,
$Q_f$ is the charge of the fermion $f$ in units of $e$, \ $N_f = 1
(3)$ for leptons (quarks), $g_{V,  f} = t_{3L, f} - 2Q_f s_W^2$
and $g_{A,  f} = t_{3L, f}$ are the vector and axial-vector
couplings of $Z$ boson to the fermion, where $t_{3L, f}$ is the
projection of the weak isospin, and
\begin{equation}
 \lambda_{f, \, W} \equiv \frac{4m_{f, \, W}^2}{q^2 }\, , \quad \lambda^{\prime}_{f, \, W} \equiv \lambda_{f, \, W} |_{q^2 = m_h^2} .
\label{eq:tau_lambda}
\end{equation}
The loop integrals for fermions, $A_f (\lambda_f^\prime,
\lambda_f)$,  and $W$ bosons,  $A_W (\lambda_W^\prime,
\lambda_W)$, are expressed in terms of the loop functions $I_{1}
(\lambda^\prime, \, \lambda)$ and  $I_{ 2} (\lambda^\prime, \,
\lambda)$ ~\cite{Spira:1998} (see~\ref{app:A}).

The coefficients $c_3, \, c_4$ in the amplitude (\ref{eq:006})
come only from the PS coupling of the Higgs boson to
fermions in the loops. We obtain
\begin{eqnarray}
c_3 &=&  \frac{1}{2} \,  \frac{g_{V,  \ell}}{q^2 - m_Z^2 + i m_Z \Gamma_Z} \,  \widetilde\Pi_{Z} + \frac{Q_{\ell}}{q^2} \, \widetilde\Pi_{\gamma } \, , \label{eq:c_3}  \\
c_4 &=&   - \frac{1}{2}  \, \frac{ g_{A,  \ell}}{q^2 - m_Z^2 + i m_Z \Gamma_Z} \, \widetilde\Pi_{ Z} \, \label{eq:c_4} ,
\end{eqnarray}
\begin{eqnarray}
\widetilde\Pi_{ Z} &=&   \frac{e g^3}{16 \pi^2m_W  } \, p_f \, \frac{2 g_{V,f}}{c_W^2} N_f Q_f  I_2 (\lambda_f^\prime, \lambda_f)  ,  \label{eq:tilde-Pi_Z}  \\
\widetilde\Pi_{ \gamma} &=&   \frac{e^3 g}{16 \pi^2 m_W} \, p_f \, 4 Q_f^2 \, N_f \, I_2 (\lambda_f^\prime, \lambda_f) \, . \label{eq:tilde-Pi_gamma}
\end{eqnarray}
Of course, the sum over all fermions $f = (\ell, q)$  in
(\ref{eq:Pi_Z}), (\ref{eq:Pi_gamma}), (\ref{eq:tilde-Pi_Z}) and
(\ref{eq:tilde-Pi_gamma}) is implied.


\subsection{Forward-backward asymmetry in the SM \label{subsec:SM}}

In the SM the angular distribution in Eq.~(\ref{eq:007}) simplifies.
Indeed, one sets $s_{\ell} = p_{\ell}=0$ in
Eq.~(\ref{eq:007}) and $s_f = p_f =0$ in Eqs.~(\ref{eq:Pi_Z}),
(\ref{eq:Pi_gamma}), (\ref{eq:tilde-Pi_Z}) and
(\ref{eq:tilde-Pi_gamma}). Then $c_{3, \, SM} = c_{4,\, SM} = 0$
and (\ref{eq:007}) turns into
\begin{eqnarray}
|{\cal M}|^2_{SM}  &=&  c_0^2 \, A   + 2 \, c_0 \, {\rm Re} (c_{1, \, SM}) \, B \, \nonumber \\
&& + |c_{1, \, SM}|^2  \, D +  |c_{2, \, SM}|^2 \, E \, ,
\label{eq:M_SM}
\end{eqnarray}
where $c_{1, \, SM} = c_{1}|_{s_f =0}$ and $c_{2, \, SM} = c_{2}|_{s_f =0}$ in (\ref{eq:c_1}) and (\ref{eq:c_2}).

In follows from Eq.~(\ref{eq:delta_Gamma}) that in the SM
\begin{equation}
A_{\rm FB } (q^2)_{SM} = 0 .
\label{eq:A_FB_SM}
\end{equation}

Therefore a nonzero value of the FB asymmetry can arise only in
certain models beyond the SM. A similar conclusion for the decay
$h \to \gamma Z \to \gamma \ell^+ \ell^-$ with on-mass-shell $Z$
boson has been inferred in~\cite{Korchin:2013,Korchin:2013jja} and is
used there to estimate the magnitude of possible $CP$ violation
effect.

The result (\ref{eq:A_FB_SM}) is at variance with the conclusion
of Refs.~\cite{Sun:2013,Akbar:2014}, where the authors have found
a nonzero FB asymmetry in the framework of the SM.
The origin of a nonzero asymmetry in Ref.~\cite{Sun:2013} is related to
the axial-vector coupling of the $Z$ boson to the fermions in the loop diagrams.

In fact, the axial-vector $Z f \bar{f}$ coupling does not contribute to the process $h \to \gamma Z^*$ (for real or virtual $Z$). This was noticed long ago in the framework of the SM in Refs.~\cite{Cahn:1979pl,Bergstrom:1985np,Barroso:1986} on the basis of the charge-conjugation parity arguments.
As an alternative argument, in~\ref{app:B} we show in the model (\ref{eq:001}) and in the SM explicit cancellation of contributions from axial-vector $Z f \bar{f}$ coupling to the fermion-loop diagrams for $h \to \gamma^* Z^*$.


\section{\label{sec:results} Results of calculations and discussion}

Let us discuss the choice of parameters $s_f$ and $p_f$ for
the Higgs coupling to the fermions in (\ref{eq:001}).
In terms of these parameters the decay width of the Higgs to fermions, except the top quark, is equal to
\begin{equation}
\Gamma (h \to f \bar{f})\, = \, \frac{N_f G_F}{4 \sqrt{2} \pi} \, m_f^2 \,
m_h \, \beta_f \bigl( |1+s_f|^2 \beta_f^2 \, + \, |p_f|^2  \bigr) \,,
 \label{eq:width_hff}
\end{equation}
where $\beta_f = \sqrt{1- 4m_f^2/m_h^2} $ is the fermion velocity
in the rest frame of $h$. Apparently, one can put  $\beta_f
\approx 1$. Then in order to keep the Higgs decay widths to
fermions equal to their SM values we impose the following
constraint on the parameters $s_f, \, p_f$
\begin{equation}
|1+s_f|^2 + |p_f|^2 = 1 .
\label{eq:s_f+p_f}
\end{equation}
In this case, in order to ascertain the exact values of the
parameters $s_f$ and $p_f$ one would need to measure polarization
characteristics of the leptons, which is not accessible at
present.

Although Eq.~(\ref{eq:s_f+p_f}) does not uniquely determine the
parameters we choose the tentative values as in Ref.~\cite{Korchin:2013}
\begin{equation}
s_f = 1/\sqrt{2}-1  ,  \quad
p_f = \pm\,1/\sqrt{2} 
\label{eq:s_f_and_p_f}
\end{equation}
for all fermions. These values imply an equal weight of $1/2$ of the S and PS couplings.

Regarding the Higgs couplings to the top quark, we will choose them by requiring that the ratios
\begin{equation}
\mu_{gg} = \frac{\Gamma(h\to gg) }{\Gamma_{\rm SM}(h\to gg)}  , \quad
\mu_{\gamma \gamma} = \frac{\Gamma(h\to \gamma \gamma) }{\Gamma_{\rm SM}(h \to \gamma \gamma)}
\label{eq:mu_g}
\end{equation}
are consistent with the recent CMS results~\cite{Khachatryan:2014ira}
\begin{equation}
\mu_{ggh, \, t \bar{t}h} = 1.13^{+0.37}_{-0.31} , \quad
\mu_{\gamma \gamma} = 1.14^{+0.26}_{-0.23} .
\label{eq:mu_g-ATLAS}
\end{equation}

This allows us to choose the following values
of parameters $s_t$ and $p_t$
\begin{equation}
s_t = -0.3  , \quad |p_t | = 0.55  .
\label{eq:s_t-p_t}
\end{equation}
With these parameters, values of $\mu_{gg}$ and $\mu_{\gamma \gamma}$ appear to be, respectively, 1.2 and 1.23.

As the interaction of the Higgs boson with the $W^\pm$ and $Z$ is not
modified compared to the interaction in the SM, the observables in the decays $h \to Z Z \to 4 \ell $ and $h \to W W \to \ell \nu_\ell \ell \nu_\ell$, \ where $\ell = (e, \, \mu )$,  are consistent with the ATLAS and CMS data and spin--parity analyses \cite{ATLAS:2013,CMS:2013}.

Numerical values of the SM parameters are taken
from~\cite{PDG:2012}, namely, the gauge boson masses, widths, and
$Z f \bar{f}$ couplings. The quark masses are chosen according
to~\cite{Heinemeyer:2013,Dittmaier:2011}, and $\sin^2 \theta_W =
1- m_W^2/m^2_Z$.

\begin{figure}[tbh]
\begin{center}
\includegraphics[width=0.42\textwidth]{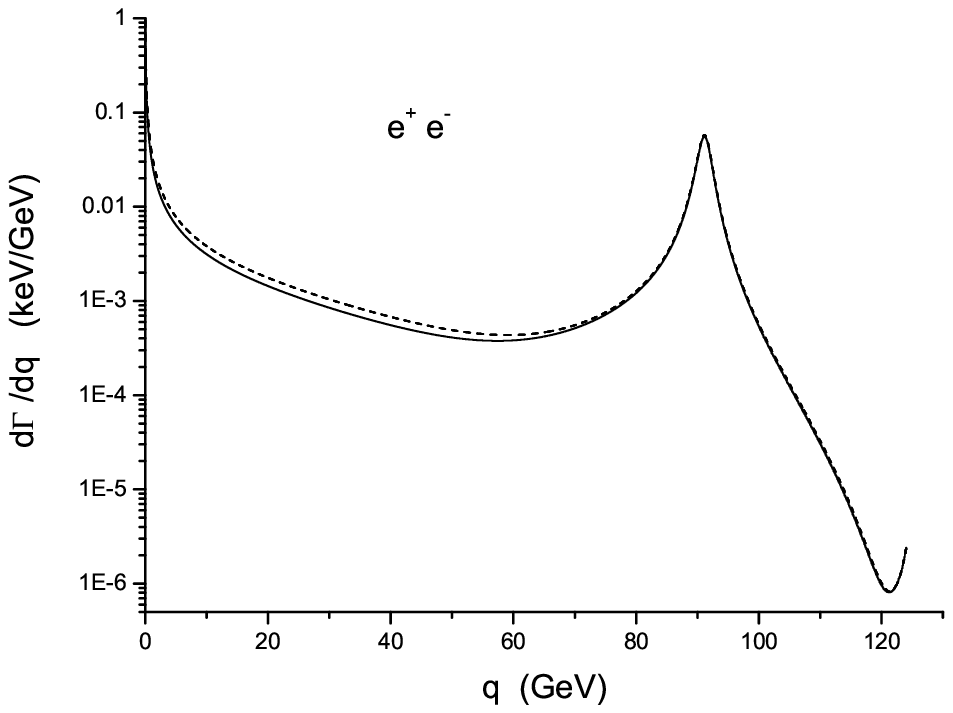}

\includegraphics[width=0.42\textwidth]{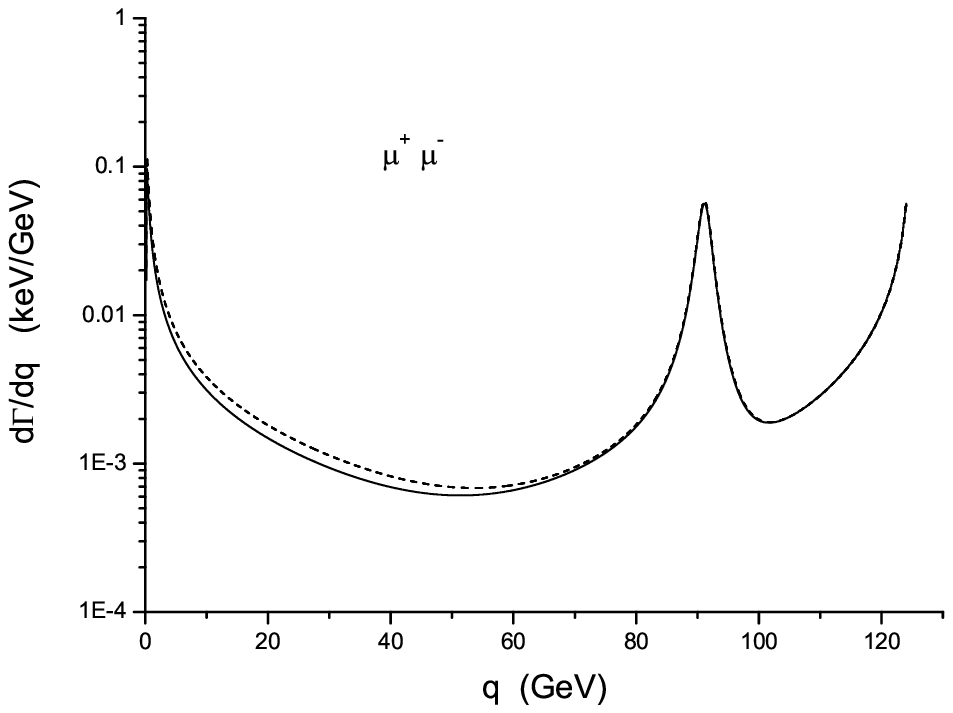}

\includegraphics[width=0.42\textwidth]{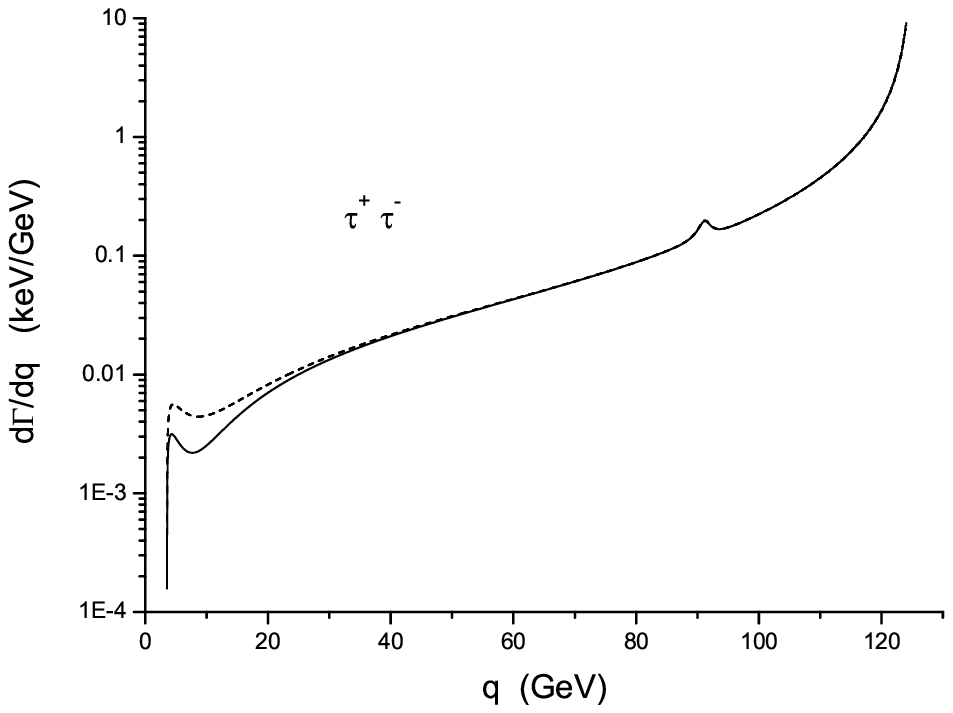}

\end{center}
\caption{Differential decay width for various final lepton pairs as a function of the dilepton invariant mass $q \equiv \sqrt{q^2}$. {\it Solid lines} are calculated in the SM, {\it dashed lines} in the model (\ref{eq:001}) (parameters NP1; see the text) }
\label{fig:widths}
\end{figure}

In Fig.~\ref{fig:widths} we show the differential decay width for
$h \to \ell^+ \ell^- \gamma$ for various leptons $\ell = (e, \, \mu,
\, \tau)$ calculated in the SM model and in the model of new
physics (\ref{eq:001}) with parameters $ s_f = 1/\sqrt{2}-1, \, p_f = + 1/\sqrt{2}$ and $s_t = -0.3, \, p_t = +0.55$. This
choice of parameters is called hereafter NP1. The photon minimal 
energy in the Higgs-boson rest frame is taken $E_\gamma = 1$
GeV in order to cut-off infrared divergence, so that $q_{max} =
(m_h^2 - 2 m_h E_\gamma)^{1/2} \approx m_h -
E_\gamma$.

As is seen from Fig.~\ref{fig:widths}, there is a deviation
from the prediction of the SM with the chosen parameters $s_f, \,
p_f$ of new physics. Integration over the invariant mass within the
interval $[q_{min}, \, q_{max}]$ leads to the widths shown in
Table~\ref{tab:widths}.
\begin{table}[tbh]
\caption{Decay width $\Gamma (h \to \gamma \ell^+ \ell^-)$ in keV
for various lepton states in the interval of invariant masses from $q_{min}$ to $q_{max}$ (in GeV) }
\begin{center}
\begin{tabular}{c c c c c}
\hline
$\ell^+ \ell^-$  & $q_{min}$  & $q_{max}$  &  SM &  NP1  \\
\hline
$e^+ e^-$   & $1$  & $124$ & $0.34$ & $0.37$  \\
$ $   & $1$  & $30$ & $0.11$ & $0.13$   \\
\hline
$\mu^+ \mu^-$     & $1$  & $124$ & $0.53$ & $0.56$  \\
$ $ & $1$ & $30$ & $0.11$ & $0.13$ \\
\hline
$\tau^+ \tau^- $ & $4$ & $124$ & $31.0$ & $31.1$ \\
$ $ & $4$ & $30$ & $0.16$ & $0.20$ \\
\hline
\end{tabular}
\end{center}
\label{tab:widths}
\end{table}

The effect of new physics appears on the level
10--20\%, if the invariant-mass interval lies below 30 GeV.
Although the decay width in this interval is very
small compared, for example, to the two-photon decay width of the
Higgs boson in the SM  $\Gamma (h \to \gamma \gamma)=9.28$ keV
(Ref.~\cite{Heinemeyer:2013}; see Table A.10 therein).

As one can also see from Fig.~\ref{fig:widths}, for the decay $h
\to \gamma e^+ e^-$ the dominant contribution to the width in
Table~\ref{tab:widths} comes from the loop amplitude. For the $h
\to \gamma \mu^+ \mu^-$ decay, the tree-level and loop
contributions are comparable, while for the $h \to \gamma \tau^+
\tau^-$ decay, the tree-level amplitude gives the dominant
contribution.

\begin{figure}[tbh]
\begin{center}
\includegraphics[width=0.42\textwidth]{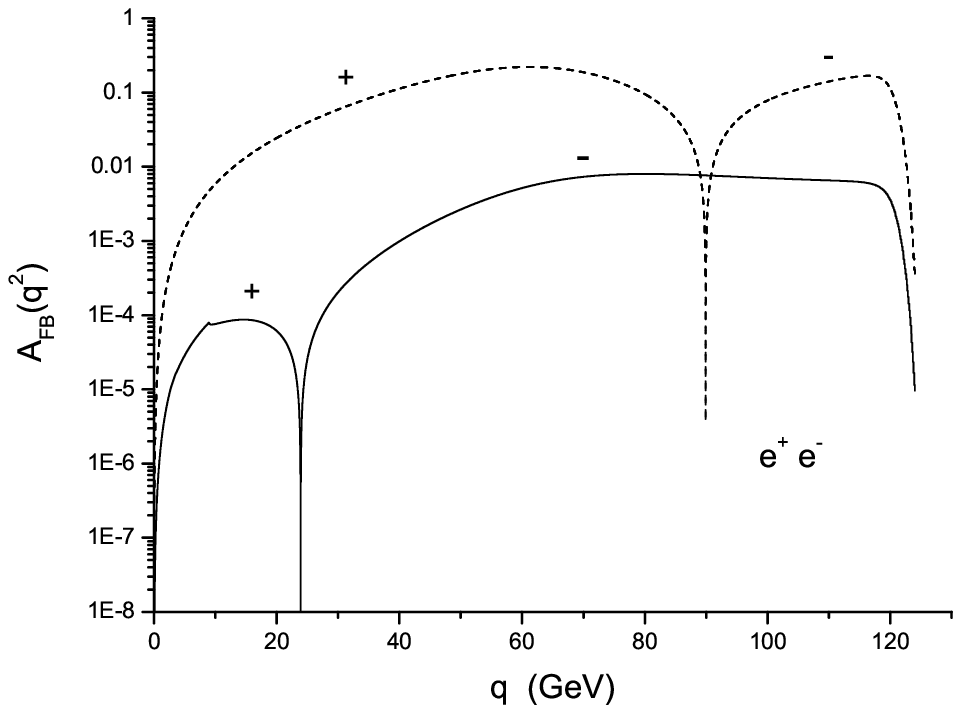}

\includegraphics[width=0.42\textwidth]{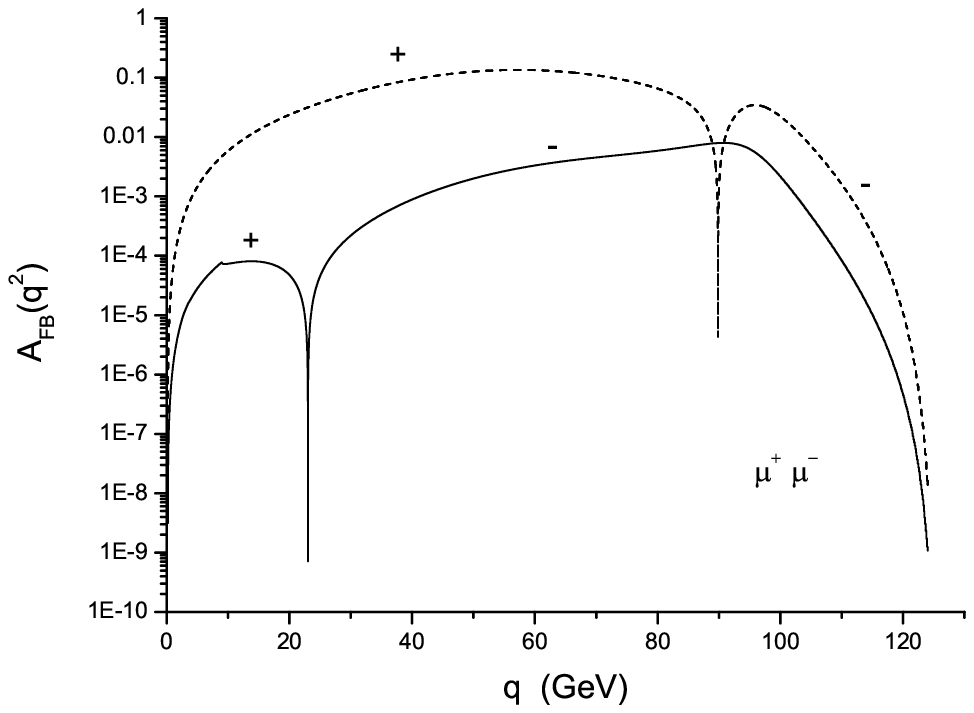}

\includegraphics[width=0.42\textwidth]{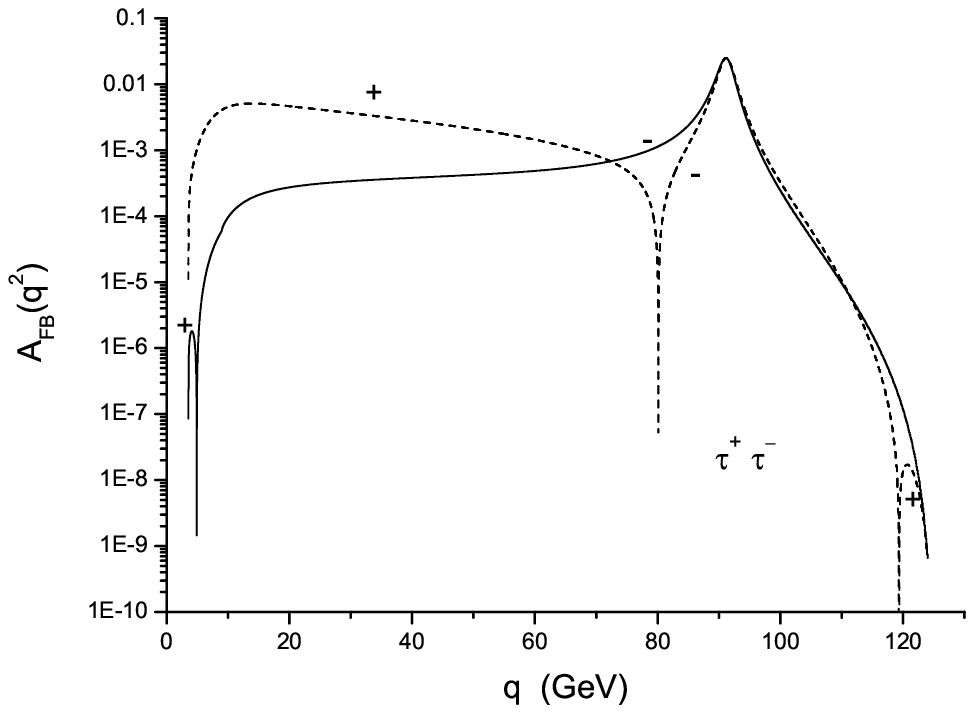}
\end{center}
\caption{Forward--backward asymmetry for various final leptons, calculated in the model (\ref{eq:001}). 
{\it Solid lines} correspond to parameters NP1 ($p_t = +0.55$),
{\it dashed lines} to parameters NP2 ($p_t = +0.55 \, i$). The sign of asymmetry is indicated at the {\it curve} }
\label{fig:asymmetries}
\end{figure}

In Fig.~\ref{fig:asymmetries} the FB asymmetry (\ref{eq:017}) is
presented as a function of $q$. As mentioned above, the FB
asymmetry can take nonzero values only in models beyond the SM, though not all models of new physics lead to nonzero FB asymmetry.
In the model (\ref{eq:001}), $A_{\rm FB}(q^2)$ is proportional to
Eq.~(\ref{eq:delta_Gamma}). All terms in (\ref{eq:delta_Gamma})
are proportional to the parameters $p_f$ which characterize PS
couplings of the Higgs to the fermions.  So in this model the FB
asymmetry is a direct measure of a possible $CP$ violation in the
$h f \bar{f}$ coupling.

For the light final leptons, $e^+ e^-$ and $\mu^+ \mu^-$,
the dominant contribution to $A_{\rm FB} (q^2)$ comes from the term in
(\ref{eq:delta_Gamma}) proportional to the imaginary part of
the combination $c_1 c_4^* + c_2 c_3^*$. This
imaginary part in turns originates from the $Z$-boson propagators in
Eqs.~(\ref{eq:Pi_Z}), (\ref{eq:Pi_gamma}), (\ref{eq:tilde-Pi_Z})
and (\ref{eq:tilde-Pi_gamma}), and the loop contributions $\Pi_Z,
\Pi_\gamma, \widetilde\Pi_{ Z}$ and $\widetilde\Pi_{ \gamma}$. The
latter have small imaginary parts arising due to the intermediate
on-mass-shell fermion--antifer\-mion pairs with the masses $m_f \le
m_h/2$. These imaginary
parts come mainly from the bottom, charm quarks, and the $\tau$ lepton
(this fact was also noticed in~\cite{Korchin:2013}).

As is seen from Fig.~\ref{fig:asymmetries}, for real values of the
parameters $s_f, \, p_f$ (see solid lines) the FB asymmetry
takes values less than $1\%$ for the electrons and muons, with
maximum value $0.8\%$ at the dilepton invariant mass around the
$Z$-boson. For the $\tau$ leptons, the FB asymmetry is bigger, with
maximum value of about $2.5\%$. In principle, observation of a nonzero FB
asymmetry will point to $CP$ violation in the Higgs coupling to
fermions, though its small values make the corresponding experimental task difficult.

Let us emphasize that real parameters $s_f, \, p_f$ follow from the
requirement of Hermiticity of the Lagrangian ${\cal L}_{h f f}$ in
Eq.~({\ref{eq:001}}).
Note that Hermiticity of Hamiltonian is a necessary condition, in
addition to Lorentz invariance, locality, and the connection
between spin and statistics, in the proof of the $CPT$ theorem in
quantum field theory \cite{Streater:1964}. It is of interest to
explore how a possible non-Hermiticity of the Lagrangian ({\ref{eq:001}}) will influence the FB asymmetry. Noticeable
sensitivity of the FB asymmetry to non-Hermiticity of the $h f
\bar{f}$ Lagrangian, in principle, can be used for testing the $CPT$ symmetry.

For the purpose of this we change for the top quark the parameter
$p_t$  from real value $0.55$ to the imaginary value $0.55\, i$,
while keeping the rest of parameters equal to their values in
model NP1. This model is hereafter called NP2. Note that this
choice of parameters does not affect the values of $\mu_{gg}$ and
$\mu_{\gamma \gamma}$ calculated above.

As a result, the FB asymmetry increases substantially, up to  22\%
for electron and 14\% for muon, while for $\tau$ lepton the
maximal value of the asymmetry remains on the level of 2.5\% (see dashed
lines in Fig.~\ref{fig:asymmetries}).

In general, $A_{\rm FB}(q^2)$ changes sign as a function of the invariant mass,
therefore the integrated FB asymmetry (\ref{eq:AFB_int}) over the whole interval of $q$ is rather small and is not a suitable observable.
In particular, in the model NP1 (NP2) the FB asymmetry integrated over the interval $[1, \; 124]$ GeV for electrons and muons is  $\langle A_{\rm FB} \rangle = -0.4\%$ ($+1\%$), and integrated over the interval $[4, \; 124]$ GeV for $\tau$ leptons is $\langle A_{\rm FB} \rangle = -0.06\%$ ($-0.04\%$).

However, the integrated asymmetry increases for an appropriately chosen interval of invariant mass, in which $A_{\rm FB}(q^2)$ does not change
sign. For example, within the interval $[37.5 , \; 75]$
GeV, for the $e^+ e^-$ pair $\langle A_{\rm FB} \rangle =
-0.4\%$ \ ($+17\%$) in the model NP1 (NP2). Within the same
interval, for the $\mu^+ \mu^-$ pair  $\langle A_{\rm FB} \rangle = -
0.3\%$ \ ($+12\%$) in the model NP1 (NP2).


\section{\label{sec:conclusions} Conclusions}

The differential decay width and forward--backward
asymmetry have been calculated for the decay of Higgs boson to the
photon and lepton--antilepton pair, $h \to \gamma \ell^+ \ell^-$,
where $\ell = (e,\, \mu, \, \tau)$. The calculations were performed in the 
framework of the SM and in a model of new physics, in which the
Higgs boson interacts with fermions via a mixture of scalar and
pseudoscalar couplings. Both the tree-level amplitudes and the
one-loop $h \to \gamma Z^* \to \gamma \ell^+ \ell^-$ and $h \to \gamma \gamma^* \to \gamma \ell^+ \ell^-$ diagrams have been included.

We noted that the FB asymmetry vanishes identically in the
SM. In models of new physics, which include effects of $CP$
violation in the $h f \bar{f}$ interaction, this asymmetry takes nonzero
values. The experimental study of the FB asymmetry is of interest in
the search for effects of new physics in the Higgs--fermion
interaction.

In numerical estimates of the decay width and FB asymmetry, the
model parameters $s_f, \, p_f$ have been chosen by requiring that the
$h \to f \bar{f}$ decay widths coincide with the widths in the SM for all
leptons and quarks, except the top quark. For the latter the
parameters $s_t, \, p_t$ were constrained from the
conditions that the rates of the $h \to \gamma \gamma$ and $h \to g g$ decays
are consistent with the CMS data~\cite{Khachatryan:2014ira}.

In the differential decay widths effects of new physics appear on
the level of 10--20\%, especially at relatively small values of
dilepton invariant mass $\lesssim 30$ GeV.

As for the FB asymmetry, it takes nonzero values; however, these values are
small. In particular, $A_{\rm FB}(q^2)$ reaches
1\% for electrons and muons and 2.5\% for $\tau$ leptons in the
region of invariant mass $q \sim m_Z$.

We have also shown that the FB asymmetry increases considerably if
the parameter $p_t$ for the pseudoscalar $h t \bar{t}$ coupling
becomes complex. Specifically, for the imaginary value $p_t = 0.55\,
i$ the asymmetry rises up to 22\% for electrons and 14\% for muons
in the region of invariant mass $q \sim 50-60$ GeV.  Hence the FB
asymmetry for the $e^+ e^-$ and $\mu^+ \mu^-$ pairs turns out to be
sensitive to the non-Hermiticity of the $h t \bar{t}$ interaction
Lagrangian. Since the requirement of Hermiticity underlies the
proof of the $CPT$ theorem, the FB asymmetry may also be used for
testing the $CPT$ symmetry.

For the $\tau$ leptons, the FB asymmetry is sensitive to
non-Hermiticity of the $h t \bar{t}$ interaction Lagrangian in the
region of relatively small $q \sim 20$ GeV, staying less than 1\%.
Whereas its maximal value 2.5\% remains the same with the real
and imaginary parameter $p_t$.

In our opinion experimental study of the differential decay width and FB asymmetry in the $h \to \gamma \ell^+ \ell^-$ decays may give
additional information on the couplings of the Higgs boson to
fermions.


\appendix

\section{Definition of loop functions \label{app:A}}

The loop functions for the fermions, $A_f(\lambda_f^\prime,
\lambda_f)$, and $W^\pm$ boson, $A_W(\lambda_W^\prime,
\lambda_W)$, are equal to
\begin{equation}\label{appendixA:004}
A_{f}(\lambda^\prime, \lambda) \, = \, I_1(\lambda^\prime, \lambda) -I_2(\lambda^\prime, \lambda)\, ,
\end{equation}
\begin{eqnarray}
\label{appendixA:003}
&& A_W(\lambda^\prime, \lambda)\, = \,
16 \bigl(1- \frac{1}{\lambda} \bigr) \, I_2(\lambda^\prime ,\lambda)
\nonumber  \\
&& + \Bigl[ \bigl(1 +\frac{2}{\lambda^\prime }\bigr)  \bigl(\frac{4}{\lambda} -1 \bigr) -\bigl (5 +\frac{2}{\lambda^\prime } \bigr) \Bigr] \, I_1(\lambda^\prime ,\lambda) \,.
\end{eqnarray}

The loop functions $I_{1,2}(\lambda^\prime, \lambda)$ are defined in Ref.~\cite{Spira:1998}:
\begin{eqnarray}
I_1(\lambda^\prime, \lambda)& = &\frac{\lambda^\prime \,\lambda}{2\,(\lambda^\prime -\lambda)} \Bigl[1 + \frac{\lambda^\prime \,\lambda}{\lambda^\prime -\lambda}
\bigl( f(\lambda^\prime )-f(\lambda) \bigr) \nonumber \\
&+&\frac{2\, \lambda^\prime }{\lambda^\prime -\lambda} \bigl( g(\lambda^\prime )-g(\lambda) \bigr) \Bigr] \,,
\label{appendixA:005}
\end{eqnarray}
\begin{equation}\label{appendixA:006}
I_2(\lambda^\prime, \lambda)=-\frac{\lambda^\prime \,\lambda}{2\,(\lambda^\prime -\lambda)} \bigl( f(\lambda^\prime )-f(\lambda) \bigr) \, ,
\end{equation}
where the functions $f(\lambda)$ and $g(\lambda)$ can be expressed as
\begin{eqnarray}\label{appendixA:007}
f(\lambda)=\left\{
\begin{array}{ll}  \displaystyle
\arcsin^2\frac{1}{\sqrt{\lambda}} & \lambda\geq 1 \\
\displaystyle -\frac{1}{4}\left( \log\frac{1+\sqrt{1-\lambda}}
{1-\sqrt{1-\lambda}}-i\pi \right)^2 \hspace{0.5cm} & \lambda<1 ,
\end{array} \right.
\end{eqnarray}
\begin{eqnarray}\label{appendixA:008}
g(\lambda)=\left\{
\begin{array}{ll}  \displaystyle
\sqrt{\lambda-1}\,\arcsin\frac{1}{\sqrt{\lambda}} & \lambda\geq 1 \\
\displaystyle \frac{\sqrt{1-\lambda}}{2}\left(
\log\frac{1+\sqrt{1-\lambda}} {1-\sqrt{1-\lambda}}-i\pi \right)
\hspace{0.5cm} & \lambda<1 .
\end{array} \right.
\end{eqnarray}


\section{\label{app:B} Fermion-loop integrals for the $h \to \gamma Z^*$ transition }

Here we show that axial-vector $Z f \bar{f}$ coupling to the loop fermions
does not contribute to the process $h \to \gamma Z^* \to
\gamma \ell^+ \ell^- $. The derivation below is similar to the
proof of Furry's theorem in quantum electrodynamics (see, for
example, \cite{Berest},  \S \ 79).

\begin{figure}[tbh]
\begin{center}
\includegraphics[width=0.48\textwidth]{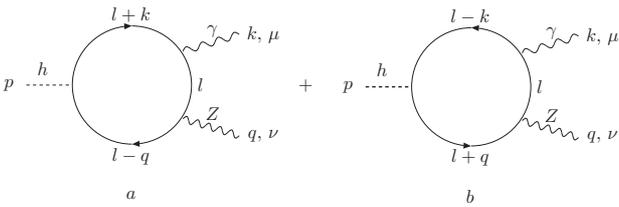}
\end{center}
\caption{Fermion-loop diagrams for the process $h \to \gamma Z $ with real/virtual $Z$ boson and photon}
\label{fig:fermions}
\end{figure}

The $h f \bar{f}$ vertex in the model (\ref{eq:001}) is proportional to the factor $(1 + s_f + i p_f \gamma_5 )$, while
the $Z f \bar{f}$ vertex is proportional to
$ \gamma^\nu (g_{V,  f} - g_{A,  f} \gamma_5)$. The diagrams `a' and `b'
in Fig.~\ref{fig:fermions} with real/virtual $Z$ and $\gamma$
correspond to the expressions (omitting irrelevant constants):
\begin{eqnarray}
T_a &=& \int {\rm d}^4 l \, {\rm Tr}  \Bigl( \gamma^\mu S(l+k) (1+s_f + i p_f \gamma_5)  \nonumber \\
& \times & S(l-q) \gamma^\nu (g_{V,  f} -g_{A,  f} \gamma_5 ) S(l) \Bigr) \, ,
\label{eq:B01}
\end{eqnarray}
\begin{eqnarray}
T_b &=& \int {\rm d}^4 l \, {\rm Tr}  \Bigl( S(l) \gamma^\nu (g_{V,  f} -g_{A,  f} \gamma_5 ) S(l+q) \nonumber \\
& \times & (1+s_f + i p_f \gamma_5) S(l-k) \gamma^\mu \Bigr) \, ,
\label{eq:B02}
\end{eqnarray}
where $S(p) = (\vslash{p} - m_f +i0)^{-1}$.

Introduce matrix $U_c$ of the charge-conjugation operator with the following properties:
\begin{eqnarray}
U_c^{-1} \, \gamma^\mu \, U_c &=&  -\gamma^{\mu \, T}\, , \quad  U_c^{-1} \, \gamma_5 \, U_c  = \gamma_5^T , \nonumber \\
U_c^{-1} \, S(p) \,  U_c & =&  S(-p)^T ,
\label{eq:B03}
\end{eqnarray}
where `$T$' means matrix transposition.

Using the unitarity conditions $U_c U_c^{-1} = U_c^{-1} U_c = 1$ we can
write for $T_b$
\begin{eqnarray}
\label{eq:B04}
 T_b &=& \int  {\rm d}^4 l \, {\rm Tr} \Bigl( S(-l)^T  \gamma^{\nu \,T} (g_{V,  f} -g_{A,  f} \gamma_5^T )   \nonumber \\
& \times &  S(-l-q)^T (1+s_f + i p_f \gamma_5^T) S(-l+k)^T \gamma^{\mu \, T} \Bigr) \, \, \nonumber\\
&=&  \int {\rm d}^4 l \, {\rm Tr}  \Bigl( \gamma^\mu S(-l+k) (1+s_f + i p_f \gamma_5)  \nonumber \\
& \times &  S(-l-q) (g_{V,  f} -g_{A,  f} \gamma_5 )
  \gamma^\nu S(-l) \Bigr)^T \,  \nonumber \\
&=&  \int {\rm d}^4 l \, {\rm Tr} \Bigl( \gamma^\mu S(l+k) (1+s_f + i p_f \gamma_5) \nonumber \\
& \times &  S(l-q) (g_{V,  f} -g_{A,  f} \gamma_5 ) \gamma^\nu S(l) \Bigr)
\nonumber \\
&=&  \int {\rm d}^4 l \, {\rm Tr}  \Bigl( \gamma^\mu S(l+k) (1+s_f + i p_f \gamma_5)  \nonumber \\
& \times & S(l-q) \gamma^\nu (g_{V,  f} + g_{A,  f} \gamma_5 ) S(l) \Bigr) \,,
 \end{eqnarray}
where the property ${\rm Tr} (A^T B^T \ldots C^T)$ = ${\rm Tr} (C
\ldots B A)$ for arbitrary matrices $A, B, C, \ldots$ is used, and  the integration variable is changed, $l \to -l$.

Adding (\ref{eq:B01}) and (\ref{eq:B04}) we obtain the sum of
diagrams `a' and `b' in Fig.~\ref{fig:fermions}:
\begin{eqnarray}
T_a +T_b &=& 2 g_{V,  f} \int {\rm d}^4 l \, {\rm Tr} \Bigl( \gamma^\mu S(l+k)
 \nonumber \\
& \times &  (1+s_f + i p_f \gamma_5) S(l-q) \gamma^\nu S(l) \Bigr)\, ,
\end{eqnarray}
which means that the contribution from the $Z f \bar{f}$ axial-vector coupling vanishes, while the contribution from the vector coupling doubles.

Setting $s_f=p_f=0$ in the Higgs fermion vertex reproduces the result in the SM.


\end{document}